# PolSeT: Polish Semantics of Timbre Dataset

JAN JASIŃSKI
*AGH University of Krakow*

This data report introduces PolSeT (Polish Semantic Timbre), a dataset designed to facilitate research in psychoacoustics and Music Information Retrieval (MIR) in Polish and cross-cultural contexts. The dataset contains data from two sequential experiments. Experiment 1 (N=60) was a free-verbalization task aimed at creating a lexicon of Polish semantic descriptors. Using 11 stimuli, a total of 1901 descriptors (701 unique) were gathered. Experiment 2 (N=105) utilized this lexicon to conduct a semantic differential study, where participants rated 18 instrument sounds on 8 bipolar scales, with repeated trials for reliability analysis. The released dataset includes raw listener responses, comprehensive demographics (experience, gender, age), audio stimuli, and extracted acoustic features with Python extraction code. This dataset addresses a gap in open timbre research data, providing both the qualitative linguistic groundwork and the quantitative ratings necessary for psychoacoustic research and the training of multilingual semantic embedding models.

## MOTIVATION

MUSICAL timbre is a multidimensional property of sound. Lacking a dedicated lexicon, its description relies predominantly on metaphorical language (e.g., bright, warm, rough) (Saitis & Weinzierl, 2019; Wallmark & Kendall, 2018; Zacharakis et al., 2015). This reliance on cross-modal metaphors has been noted across multiple languages and cultures. While studies into this topic have been conducted by researchers for English (Howard et al., 2007; Reymore, 2020; Von Bismarck, 1974), Greek (Zacharakis et al., 2014), German (Von Bismarck, 1974), French (Faure et al., 1996), and Czech (Moravec & Štěpánek, 2003) among others, very few have made their data available, often publishing only averages and resulting conclusions. Granular data containing raw listener answers is limited and even dedicated timbre evaluation datasets typically contain only averaged results (Jiang et al., 2019). This limitation hinders the development of cross-cultural models of timbre perception and culturally aware Music Information Retrieval (MIR) systems.

The PolSeT dataset was created to facilitate research into the semantics of timbre description in the Polish language and to enable the inclusion of Polish in cross-cultural analyses. Its raw, row-level ratings offer researchers insight into demographic effects and mappings between acoustic features and semantic labels, while also providing data for the training and testing of music timbre perception models.

Subsets of this dataset have been previously analyzed in published studies examining Polish timbre semantics from linguistic (Jasiński & Markowski, 2025) and psychoacoustic (Markowski et al., 2022) perspectives. Those publications reported aggregated results and domain-specific analyses without releasing the underlying raw data. This data report addresses that gap by providing the complete dataset with comprehensive documentation, enabling further use, research and reproducibility studies.

## METHODS

The PolSeT dataset is divided into two sections between two experiments conducted in 2021. The first one aimed to create a lexicon of words used in the Polish language to describe the timbre of musical instruments. To this end a free-verbalization task was performed. The second experiment was a semantic differential task. The resulting data contains connections between audio and listener perception of audio as described on semantic scales. This section contains a detailed description of both experiments.

### Experiment 1

The goal of this study was to present listeners with varied audio stimuli and gather as many words and phrases as they would use to describe the timbre of the sound. The experiment was conducted remotely through the internet with participants using headphones. This approach was considered acceptable as the audio stimuli were only meant as inspiration for the free-verbalization task. The participants were asked to enter their age, gender, musical experience and the number of hours they spend each week listening to music. Professional experience was defined as completion of a music school degree, higher education in musicology, acoustic engineering or related fields, or professional work in the music industry. Amateur experience was defined as playing an instrument or producing music as a hobby for more than one year.

Participants were presented with the same interface for each sound (see Figure 1). To avoid influencing responses through audio stimuli order it was randomly generated for each participant. Participants could listen to the stimulus repeatedly. To proceed, they were required to input at least one descriptor before clicking 'Next'.

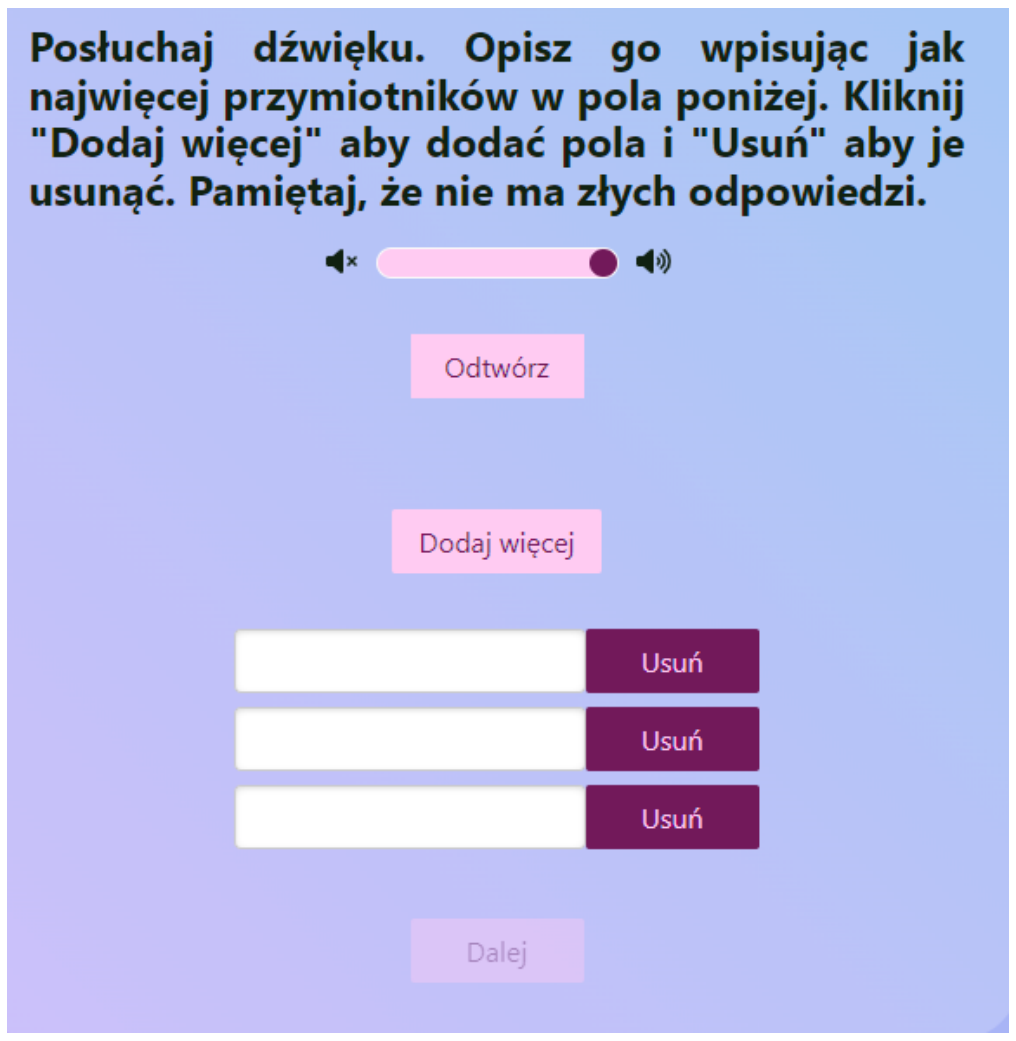

**Fig. 1.** The graphical user interface used in Experiment 1 (Free Verbalization). The Polish instructions translate to: Listen to the sound. Describe it using as many adjectives using the text boxes. Click "Add more" or "Delete" to add and remove boxes. Remember, there are no wrong answers

The stimulus set for experiment 1 consisted of 11 diverse sounds selected to elicit a wide range of descriptors. These consisted of a single note played by synthesizers and instruments from different cultures chosen to maximize audio feature variance. The audio files are part of the published dataset.

### Experiment 2

The goal of this study was to examine connections between the acoustic features and semantic descriptions in the Polish language of instrument sounds. To this end a semantic differential experiment was conducted utilizing a controlled field listening setup consisting of Beyerdynamic DT 770 Pro closed back headphones, a M-audio Fasttrack Pro audio interface used as a headphone amplifier and a laptop running the created experimental software. Such a solution, utilized in quiet environments, provided more flexibility than traditional, stationary psychoacoustic setups, while maintaining a high level of consistency with regards to the perceptual stimuli the participants were experiencing and describing.

Eight semantic scales were selected for this experiment based on the most frequent antonym pairs from Experiment 1. Priority was given to scales that align with previous research in other languages (Alluri & Toiviainen, 2010; Faure et al., 1996; Von Bismarck, 1974; Zacharakis et al., 2014) to facilitate cross-cultural comparison. The final scales are presented in table 1 together with their English translations.

**Tab. 1.** Semantic scales used in experiment 2 together with the authors' translation into English.

| Polish scales | | English translation | |
|---|---|---|---|
| Ciemny | Jasny | Dark | Bright |
| Zabrudzony | Czysty | Dirty | Clean |
| Ciepły | Zimny | Warm | Cold |
| Dźwięczny | Głuchy | Resonant | Dull |
| Ostry | Łagodny | Sharp | Gentle |
| Miękki | Twardy | Soft | Hard |
| Wybrzmiewający | Stłumiony | Ringing | Muted |
| Syntetyczny | Naturalny | Synthetic | Natural |

The study utilized 18 distinct audio stimuli. Sound selection was guided by four key criteria: organological diversity, comparative structure (creating distinct subsets for pairwise comparison), cultural breadth and acoustic variance. The final set includes standard orchestral instruments (e.g. Violin, Clarinet, Cello), contemporary sources (e.g. Fuzz Guitar, FM Synthesizer), and non-Western instruments (e.g. Sitar, Koto). Stimuli were generated using high-fidelity digital synthesis and sampling rather than live recordings. This approach ensured strict control and loudness normalization (-24.5 LUFS). Most stimuli were rendered at C4 (~261.6 Hz), with minor register adjustments for selected instruments to preserve characteristic timbral features (Pipe Organ: C3 (~130.8 Hz); Marimba and Vibraphone: C5 (~523.3 Hz)).

Participants first provided demographic information (age, gender, musical experience, and weekly listening time). After setting a comfortable listening level, they rated each stimulus on eight bipolar semantic scales (Tab. 1) using a custom interface (Figure 2). Stimulus order, scale order and scale polarity were randomized for each participant. To enable assessment of intra-rater reliability, five stimuli were presented twice without participants' awareness.

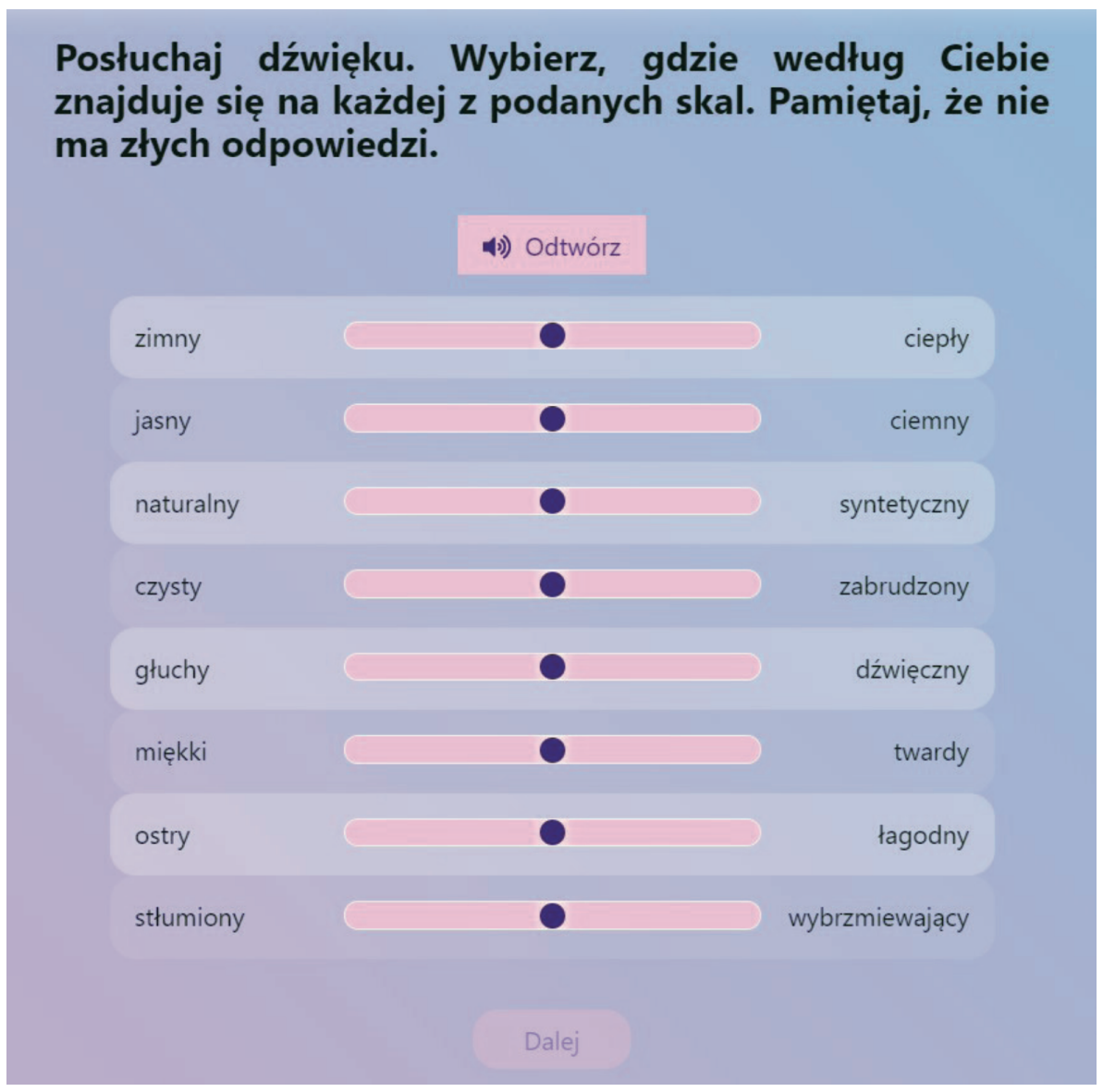

**Fig. 2.** The graphical user interface used in Experiment 2 (Semantic Differential). The Polish instructions translate to: Listen to the sound. Chose where, according to you, it lands on each of the scales. Remember, there are no wrong answers.

## DATA

### Experiment 1

The dataset for Experiment 1 includes responses from 60 participants. The demographic distribution comprises 32 males, 27 females, and 1 non-binary individual. Age distribution is skewed heavily toward younger demographics (Figure 3), the 18-24 age group constitutes the largest cohort (n=31), followed by the 25-34 group (n=16), with participation declining in older brackets. As detailed in Figure 3, the majority of participants identified as having 'Amateur' musical experience (n=31) or 'None' (n=24), with a smaller subset of 'Professionals' (n=5). The free-verbalization task yielded a total of 1,901 descriptors. Following basic text normalization (lowercasing and whitespace trimming), the final lexicon contains 701 unique Polish words or phrases. Figure 4 illustrates the 20 most frequent descriptors, with strunowy (string-like, n=46) and nieprzyjemny (unpleasant, n=41) appearing most often.

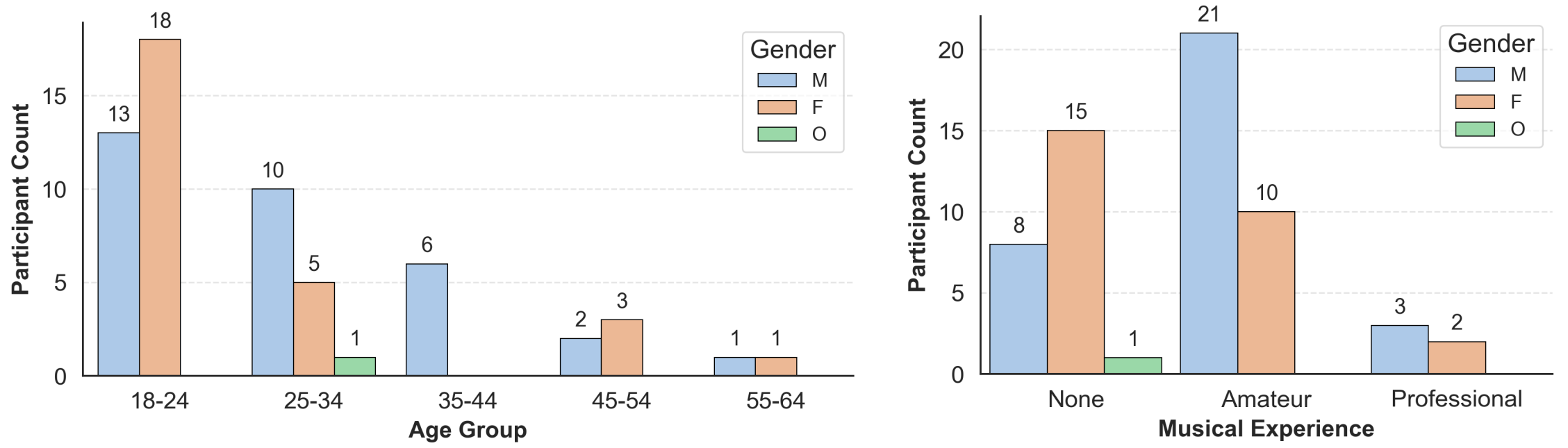

**Fig. 3.** Demographic distribution of participants in Experiment 1. (Left) Participant counts grouped by age bracket and gender. (Right) Participant counts grouped by self-reported musical experience and gender.

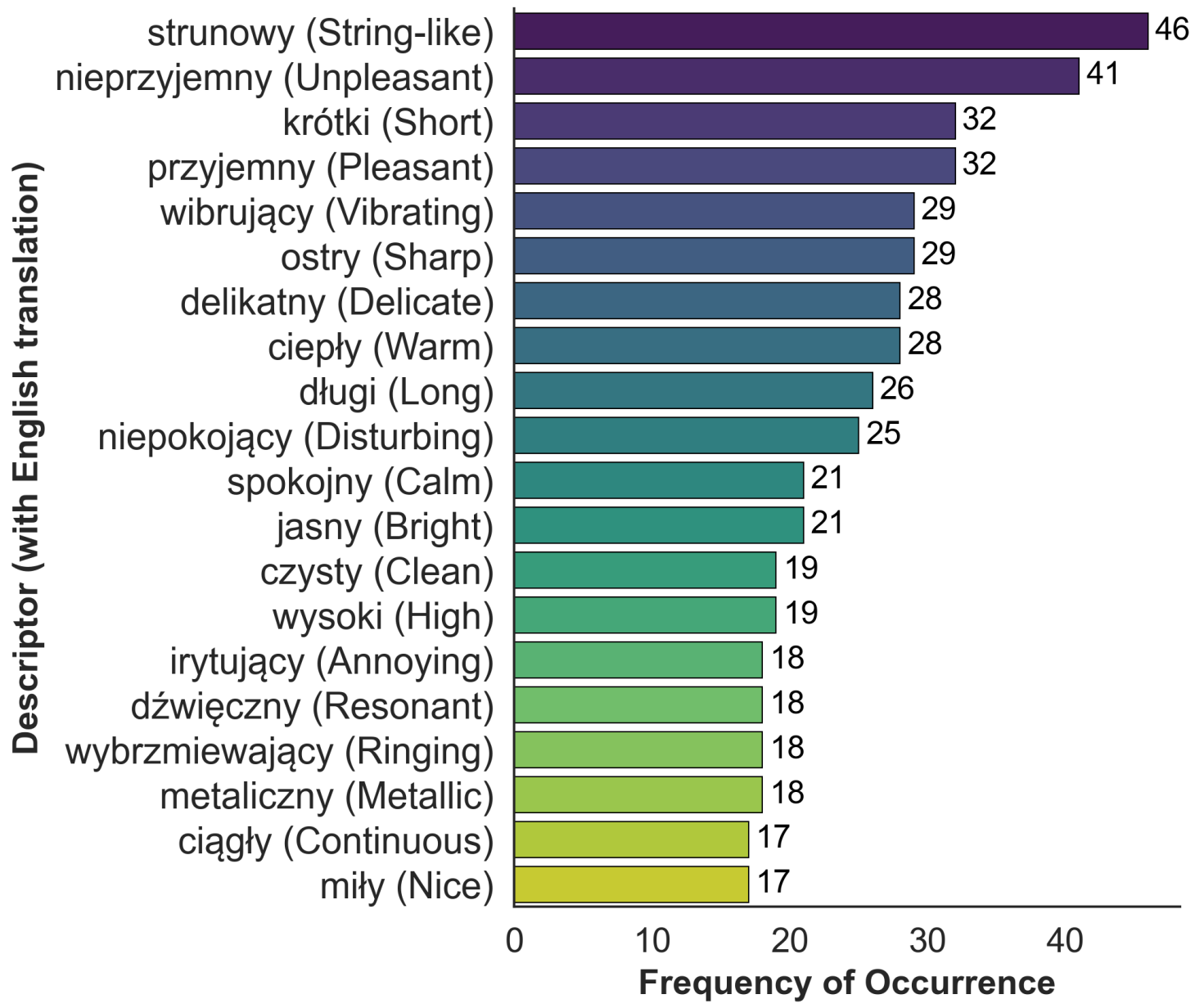

**Fig. 4.** The 20 most frequently occurring semantic descriptors from Experiment 1. Polish terms are presented with their English translations. The count represents the number of times a descriptor was used across all 11 stimuli by all participants (N=60).

### Experiment 2

A total of 105 participants completed Experiment 2, generating 19,320 individual ratings across 18 sounds and 5 repetitions. The gender distribution comprised 49 males, 53 females, and 3 non-binary/other individuals. Consistent with the previous experiment, the age demographic was heavily skewed toward younger listeners. The 18–24 age group constituted the majority (n=64), followed by the 25–34 group (n=40), with only a single participant falling into the 35–44 bracket. Musical experience was relatively evenly distributed, containing 41 participants with no experience, 31 amateurs, and 33 professionals. Figure 5 visualizes these distributions.

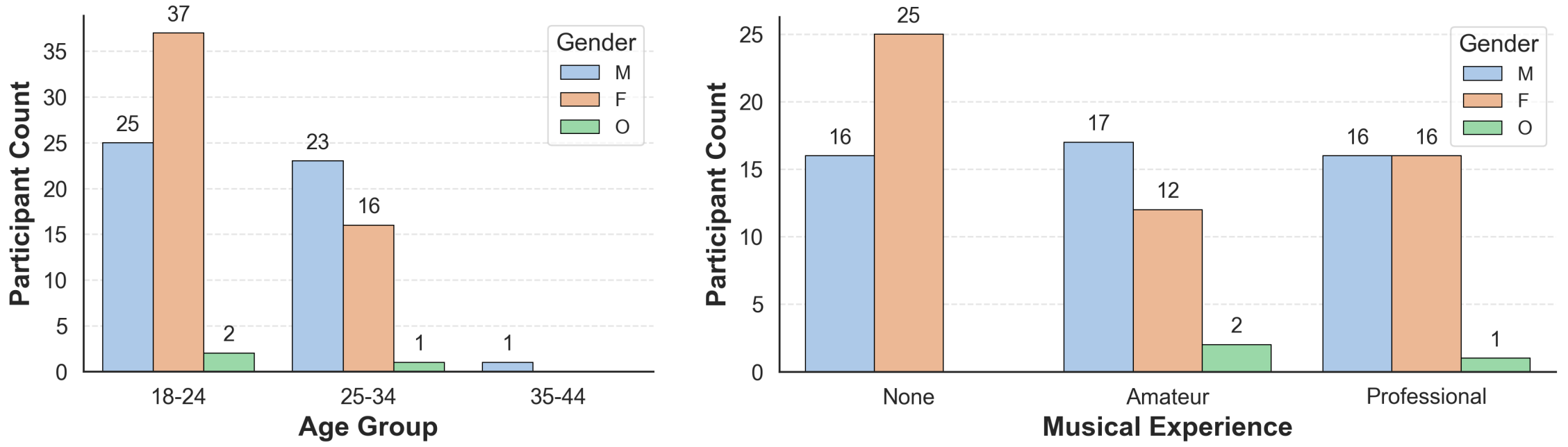

**Fig. 5.** Demographic distribution of participants in Experiment 2. (Left) Participant counts grouped by age bracket and gender. (Right) Participant counts grouped by self-reported musical experience and gender.

DATA VALIDATION

Intra-rater reliability analysis confirmed the robustness of the collected ratings. The cohort demonstrated strong consistency, with a mean Pearson correlation of r = 0.65 between repeated trials. This was corroborated by a mean Quadratic Weighted Kappa of κ = 0.63. Figure 6 presents the counts of Pearson correlations calculated between the original and repeated ratings for each participant. Most respondents achieve a strong correlation result with only 8 falling below a 0.4 threshold. Correlation coefficients between semantic scales are provided to characterize inter-scale relationships within the dataset.

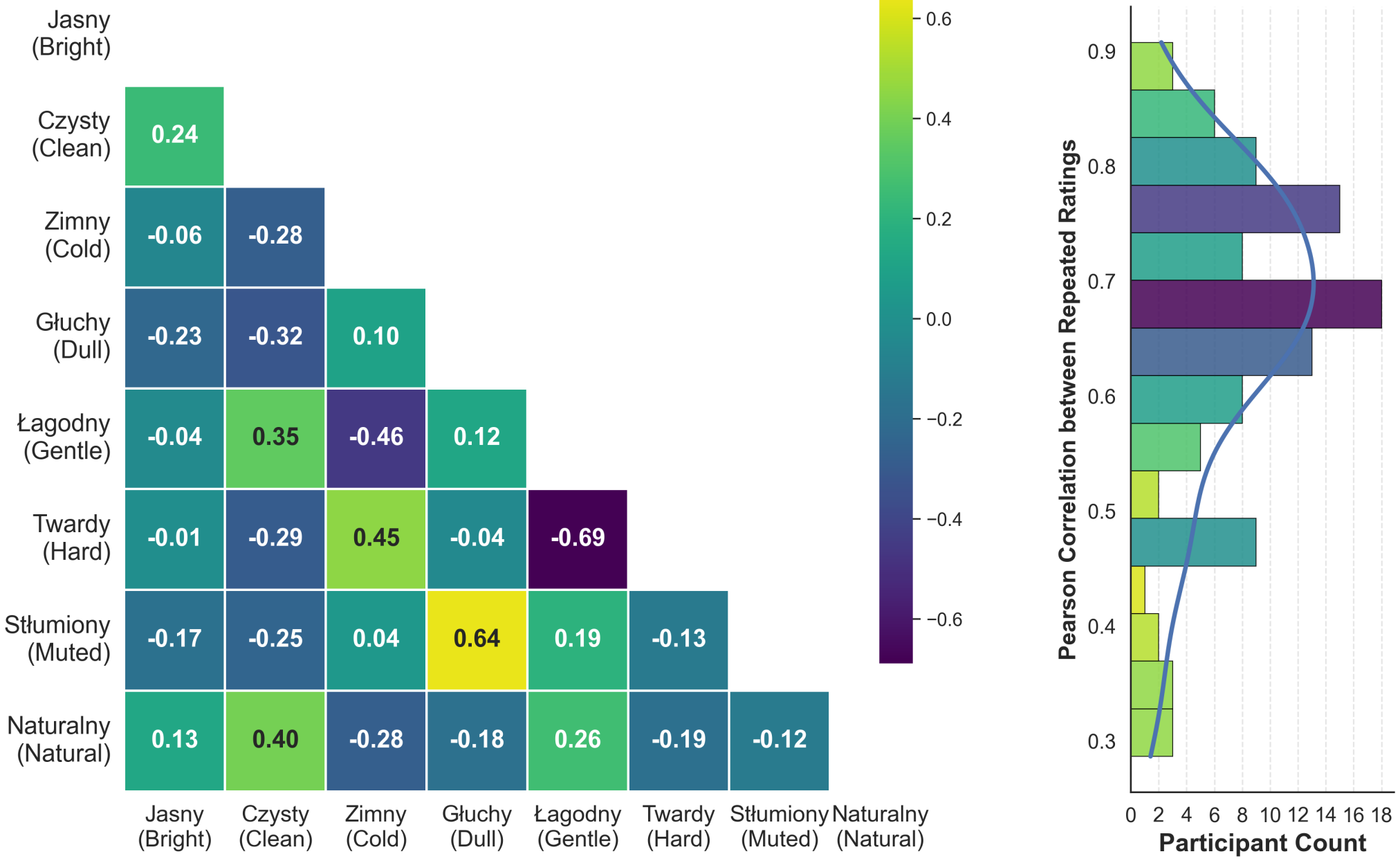

**Fig. 6.** Data validation analysis for Experiment 2. (Left) Correlation matrix of the 8 semantic scales across all ratings. (Right) Histogram of intra-rater reliability, displaying the distribution of Pearson correlation coefficients calculated between original and repeated trials for each participant.

ACOUSTIC FEATURES

A set of acoustic features was extracted from the audio files used in Experiment 2, including spectral descriptors (centroid, spread, rolloff, skewness, kurtosis, flatness), temporal measures (zero-crossing rate, spectral flux), energy-related measures (RMS), harmonic features (harmonic-to-noise ratio, tristimulus 1–3), and MFCCs (1–13) (McFee et al., 2015; Peeters et al., 2011). For each feature, both mean and standard deviation values are provided in a CSV file. Figure 7 presents a dual-projection Principal Component Analysis (PCA) of the acoustic feature space providing an overview of the dataset's acoustic variability.

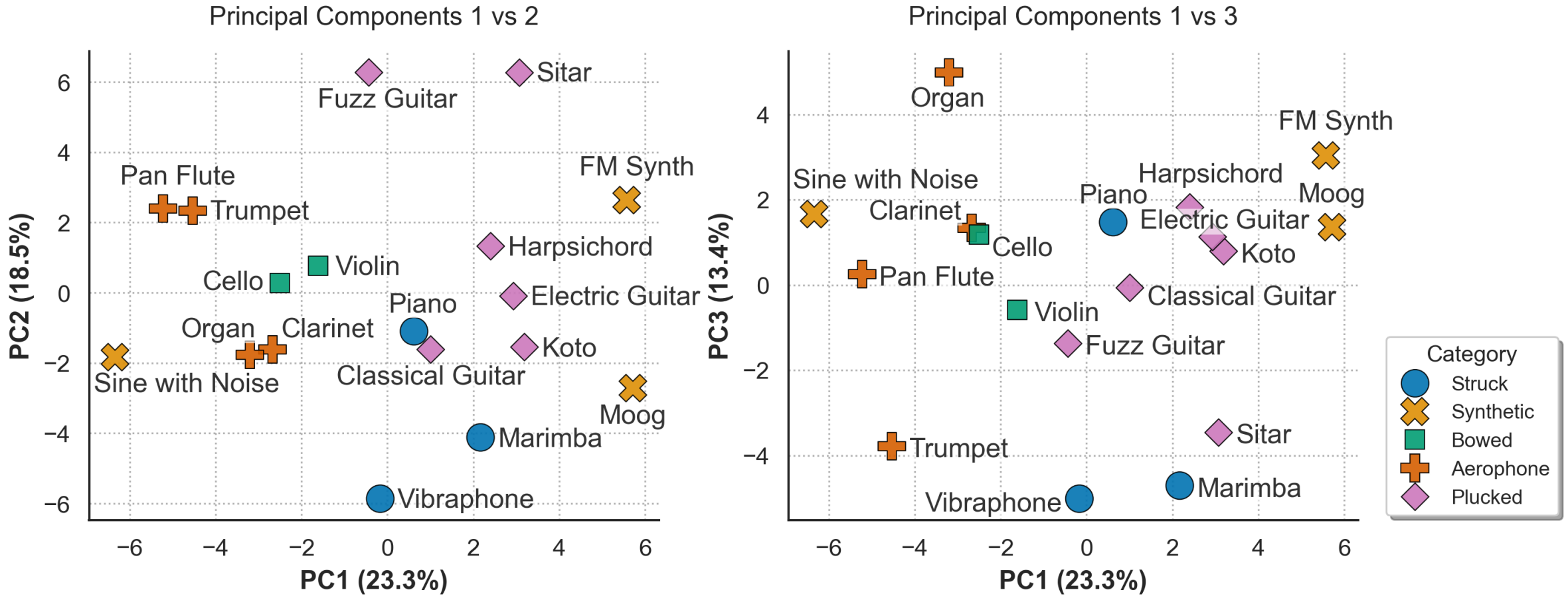

**Fig. 7.** Dual-projection Principal Component Analysis (PCA) of the acoustic feature space. (Left) The projection of PC1 vs. PC2 explains 41.8% of the data variance (Right) The projection of PC1 vs. PC3 adds a third dimension cumulatively explaining 55.2% of the variance.

## TABLE OF SPECIFICATION

The structure of the dataset, including formats and content descriptions, is summarized in Table 2.

**Tab. 2.** Overview of the PolSeT dataset structure, including file names, formats, and detailed content descriptions for each experimental component.

| File / Directory | Format | Content Description |
|---|---|---|
| /experiment_1_lexicon | | |
| polset_exp1_lexicon.csv | CSV | Single-word descriptors linked to stimuli and demographics |
| polset_exp1_participants.csv | CSV | Participant information: age, gender, musical experience and listening habits |
| polset_exp1_raw_data.json | JSON | Raw export of the full experimental data |
| /experiment_2_ratings | | |
| polset_exp2_ratings.csv | CSV | Rating values for 23 (18 + 5) stimuli across 8 bipolar semantic scales from 105 participants |
| polset_exp2_participants.csv | CSV | Participant information: age, gender, musical experience and listening habits |
| polset_exp2_scales_glossary.csv | CSV | Polish-English translations for all scale anchors |
| polset_exp2_raw_responses.zip | ZIP | Raw export of the full experimental data. |
| /audio_features | | |
| polset_exp2_acoustic_features.csv | CSV | Extracted audio feature values |
| calculate_acoustic_features.py | PY | Python script used for feature extraction |
| /audio | | |
| /exp1_stimuli/* | WAV | 11 stimuli used in experiment 1 |
| /exp2_stimuli/* | WAV | 18 stimuli used in experiment 2 |
| README.md | MD | Detailed description of each file and its format |

## DATASET USAGE

The PolSeT dataset, consisting of all raw experimental data, converted response data for easy use, participant data, audio files, audio features and documentation can be found at https://zenodo.org/records/17830609 (Jasiński et al., 2025). All data files are provided in open formats (CSV, JSON, WAV, ZIP). All participants provided informed consent prior to participation. The dataset contains no personally identifiable information and is released under a CC BY 4.0 license. This dataset is suitable for conducting Polish timbre semantic description analysis, conducting comparative linguistic studies between Polish, English and other language timbre semantics, development of cross-cultural music information retrieval algorithms and semantic embeddings and testing timbre classification models.

## ACKNOWLEDGMENTS

The author acknowledges Miron Markowski and Urszula Kaleta for their valuable contribution to the data acquisition process.

## NOTES

[1] Correspondence can be addressed to: jjasinsk@agh.edu.pl